\documentclass[10pt,twocolumn,floatfix,superscriptaddress,preprintnumbers,amsmath,amssymb,prb]{revtex4-2}
\usepackage{graphicx,bm}
\usepackage{array}
\usepackage{amssymb,amsmath}
\usepackage{float}
\usepackage{epstopdf}
\usepackage[dvips]{color}

\newcolumntype{C}{>{\centering\arraybackslash}m{1.9cm}}
\newcolumntype{L}{>{\centering\arraybackslash}m{2.75cm}}
\newcolumntype{P}{>{\centering\arraybackslash}m{1.25cm}}

\begin{document}

\title{Evidence for the coexistence of time-reversal symmetry breaking and Bardeen-Cooper-Schrieffer-like superconductivity in La$_{7}$Pd$_{3}$}

\author{D. A. Mayoh}
\email[]{d.mayoh.1@warwick.ac.uk}
\affiliation{Physics Department, University of Warwick, Coventry, CV4 7AL, United Kingdom}

\author{A. D. Hillier}
\affiliation{ISIS Facility, STFC Rutherford Appleton Laboratory, Harwell Science and Innovation Campus, Oxfordshire OX11 0QX, United Kingdom}

\author{G. Balakrishnan}
\affiliation{Physics Department, University of Warwick, Coventry, CV4 7AL, United Kingdom}

\author{M. R. Lees}
\email[]{m.r.lees@warwick.ac.uk}
\affiliation{Physics Department, University of Warwick, Coventry, CV4 7AL, United Kingdom}

\begin{abstract}
Time-reversal symmetry breaking (TRSB) with a Bardeen-Cooper-Schrieffer (BCS) -like superconductivity occurs in a small, but growing number of noncentrosymmetric (NCS) materials, although the mechanism is poorly understood. We present heat capacity, magnetization, resistivity, and muon spin resonance/relaxation ($\mu$SR) measurements on polycrystalline samples of NCS La$_{7}$Pd$_{3}$. Transverse-field $\mu$SR and heat capacity data show La$_{7}$Pd$_{3}$ is a type-II superconductor with a BCS-like gap structure, while zero-field $\mu$SR results provide evidence of TRSB. We discuss the implications of these results for both the La$_{7}X_{3}$ (where $X=$~Ni, Pd, Rh, Ir) group of superconductors and other CS and NCS superconductors for which TRSB has been observed.
\end{abstract}

\maketitle
\section{Introduction}

The discovery of time-reversal symmetry breaking (TRSB) in polycrystalline La$_{7}$Ir$_{3}$ has provided a new group of superconductors in which to investigate unconventional superconducting behavior~\cite{La7Ir3}. The report of an isotropic $s$-wave gap symmetry (nodeless) superconductivity in La$_{7}$Ir$_{3}$, along with density functional theory calculations suggesting that the enhanced specific heat in this compound is due to electron-phonon coupling, have raised questions about the origin of the TRSB in this material~\cite{La7Ir3, Li17}. Time-reversal symmetry breaking has also been observed in La$_{7}$Rh$_{3}$~\cite{Singh20} suggesting this is a common feature of the La$_{7}X_{3}$ (where $X =$~Ni, Pd, Rh, Ir) series. Here we present evidence of time-reversal symmetry breaking in La$_{7}$Pd$_{3}$ which further cements this claim.

Noncentrosymmetric (NCS) superconductors have garnered considerable interest in recent years. Their lack of a center of inversion means that parity is no longer a good quantum number~\cite{Bauer12}. A Rashba-type anti-symmetric spin-orbit coupling is allowed, lifting the degeneracy of the Fermi surface. The superconducting Cooper pairs may then form with a mixture of spin-singlet or spin-triplet components~\cite{Sigrist09, Sigrist07}. The superconducting gap function $\Delta$ can be described by $\Delta = i\tilde{\sigma}_y\left[\psi\left(\bm{k}\right)+\bm{d}\left(\bm{k}\right)\cdot\bm{\tilde{\sigma}}\right]$, where for a pure singlet $\bm{d}\left(\bm{k}\right) = 0$ and the triplet case corresponds to $\psi\left(\bm{k}\right) = 0$. It is important to note that if the triplet term is small, there will be a nodeless, near isotropic gap, making it experimentally difficult to distinguish from a Bardeen-Cooper-Schrieffer (BCS) -like $s$-wave superconductor.

A noncentrosymmetric superconductor with a mixture of triplet and singlet states may be expected to display unconventional superconducting properties. A large number of NCS superconductors have now been studied and many have indeed been shown to exhibit exotic superconducting behavior~\cite{Smidman17,Bauer12}. For example, CePt$_{3}$Si, is an antiferromagnetic heavy fermion superconductor~\cite{Bauer04}, BiPd, has an unconventional order parameter~\cite{Sun15}, Li$_{2}$Pt$_{3}$B, has a line-nodal gap structure~\cite{yuan2006,nishiyama2007,Takeya2007,Harada2012}, and (Ta/Nb)Rh$_{2}$B$_{2}$, are multigap superconductors with upper critical fields that violate the Pauli limit~\cite{Carnicom18,Mayoh18,Mayoh19}.

The possibility of spin-triplet superconductivity or singlet-triplet mixing make NCS superconductors prime candidates to exhibit time-reversal symmetry breaking. Muon spectroscopy studies of several NCS superconductors, including LaNiC$_2$~\cite{LaNiC2} and several members of the Re$_{6}T$ ($T=$~Zr, Hf, Ti) $\alpha$-Mn group of materials~\cite{Re6Zr,Re6Hfc,Shang18a,Re6Ti,Shang18} have confirmed the presence of spontaneous magnetic moments that arise below the superconducting transition temperature, $T_\mathrm{c}$, when time-reversal symmetry is broken~\cite{Ghosh20}. However, TRSB has also been observed in a number of centrosymmetric superconductors such as Sr$_{2}$RuO$_{4}$~\cite{Sr2RuO4,OpticalSr2RuO4}, PrPt$_{4}$Ge$_{12}$~\cite{PrPt4Ge12}, (Pr,La)(Os, Ru)$_{4}$Sb$_{12}$~\cite{PrOsSb,PrOsRuSb}, LaNiGa$_{2}$~\cite{LaNiGa2}, UPt$_{3}$~\cite{UPt3,OneUPt3,TwoUPt3} and (U,Th)Be$_{13}$~\cite{UThBe}, and Lu$_{5}$Rh$_{6}$Sn$_{18}$~\cite{Lu5Rh6Sn18} and even elemental rhenium~\cite{Shang18a}, while for a large number of NCS superconductors, time-reversal symmetry is preserved~\cite{Smidman17}. This leaves open the important question of how the occurrence of time-reversal symmetry breaking, the crystallographic structure (NCS or CS), and the nature of the superconducting pairing mechanism are related to one another.

In this paper we discuss the properties of La$_{7}$Pd$_{3}$ which is one of a group of hexagonal NCS superconductors with a Th$_7$Fe$_3$-type structure (space group $P6_{3}mc$). By investigating the superconducting ground state of La$_{7}$Pd$_{3}$ using muon spectroscopy further insight can be gained into the unusual superconducting behavior of this group of intermetallic compounds. Pedrazzini \textit{et al}. previously reported some of the superconducting and normal-state properties of La$_{7}$Pd$_{3}$ as well as other members of this group of superconductors, however, the fact that this family of materials has a noncentrosymmetric structure was not emphasized~\cite{Pedrazzini00}. Here, the superconducting and normal-state properties of La$_{7}$Pd$_{3}$ are investigated using magnetization, heat capacity, and resistivity measurements. Transverse-field muon spin rotation data are presented that indicate La$_{7}$Pd$_{3}$ has a conventional $s$-wave superconducting gap symmetry. Zero-field muon spin relaxation curves provide evidence of time-reversal symmetry breaking in La$_{7}$Pd$_{3}$. Finally, we compare our results on La$_{7}$Pd$_{3}$ with those obtained for other La$_{7}X_{3}$ superconductors and the Re-based $\alpha$-Mn superconductors.

\section{Experimental Details}

Polycrystalline samples of La$_{7}$Pd$_{3}$ were prepared from stoichiometric quantities of La (3N) and Pd (3N) in an arc furnace under an argon atmosphere on a water-cooled copper hearth. The sample buttons were melted and flipped several times to ensure phase homogeneity. The observed weight loss during the melting was negligible. The sample buttons were then sealed in an evacuated quartz tube, and annealed for 5 days at $500~^{\circ}$C. The material is air sensitive and was observed to rapidly develop an orange surface discoloration if exposed to air. The samples were stored in a glove-box under an argon atmosphere. A Quantum Design Physical Property Measurement System was used to measure both the heat capacity and electrical resistivity between 1.8 and 300~K in applied fields up to 9~T. A Quantum Design He-3 insert was used to access temperatures down to 0.5~K. A Quantum Design Magnetic Property Measurement System with iQuantum He-3 insert was used to measure the magnetization between 0.5 and 300~K in applied fields up to 7~T. Muon spin relaxation/rotation ($\mu$SR) measurements where performed using the MuSR spectrometer at the ISIS Neutron and Muon Source at the Rutherford Appleton Laboratory, UK~
$\mu$SR measurements were performed in both transverse-field (TF) and zero-field (ZF) modes. A full description of the detector geometries can be found in Ref.~\onlinecite{Lee96}. A crushed sample of La$_{7}$Pd$_{3}$ was mounted on a 99.995\% silver plate and inserted into a dilution refrigerator to measure at temperatures from 50~mK to 4~K.

\section{Characterization}
\subsection{Magnetization and electrical resistivity measurements}
\label{MagRes}

\begin{figure}[p]
\centering
\includegraphics[width=0.8\columnwidth]{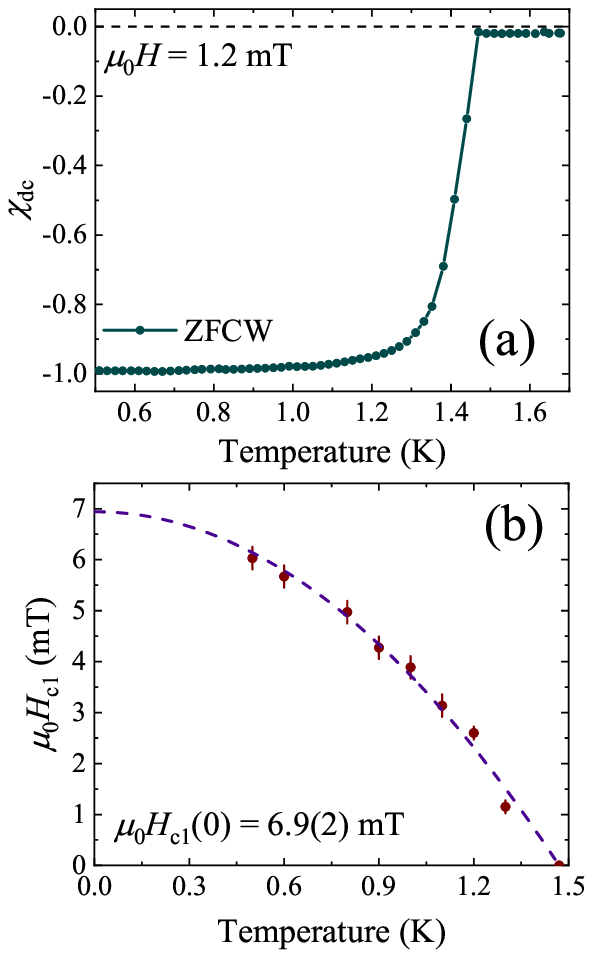}
\caption{(a) Temperature dependence of the dc volume magnetic susceptibility $\chi_{\mathrm{dc}}$ for La$_{7}$Pd$_{3}$, in zero-field-cooled warming (ZFCW) mode, measured in an applied magnetic field 1.2~mT showing a superconducting onset temperature $T^{\mathrm{onset}}_{\mathrm{c}} = 1.46(5)$~K. (b) Lower critical field $H_{\mathrm{c1}}$ versus temperature for La$_{7}$Pd$_{3}$. The dashed line shows a fit to the data using $H_{\mathrm{c1}}\left(T\right) = H_{\mathrm{c1}}\left(0\right)\left[1 - t_{\mathrm{r}}^{2}\right]$ which gives $\mu_{0}H_{\mathrm{c1}}\left(0\right) = 6.9(2)$~ mT.}
\label{FIG: La7Pd3 Magnetization}
\end{figure}

\begin{figure}[p]
\centering
\includegraphics[width=0.8\columnwidth]{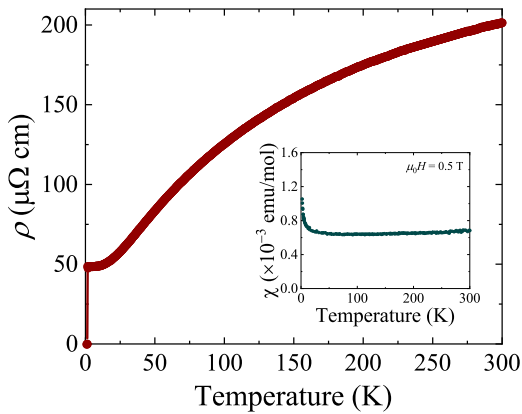}
\caption{Normal state properties of La$_{7}$Pd$_{3}$ : Temperature dependence of electronic resistivity from 1 to 300~K with zero applied magnetic field. The residual resistivity ratio (RRR) for La$_{7}$Pd$_{3}$ is approximately 4.2 and the residual resistivity just above the transition, $\rho_{0}\left(\mathrm{2~K}\right) = 48.4(3)~\mu\Omega$~cm. Inset: Temperature dependence of the magnetic susceptibility in an applied field of 0.5~T (in the normal state). An upturn at low temperature is consistent with the small quantity of paramagnetic impurities present in the La used to prepare the sample.}
\label{FIG: La7Pd3 Resistivity}
\end{figure}

To confirm bulk superconductivity in the La$_{7}$Pd$_{3}$ samples the dc volume magnetic susceptibility as a function of temperature, $\chi_{\mathrm{dc}}\left(T\right)$, was measured in an applied field of 1.2~mT between 0.5 and 1.7~K as shown in Fig.~\ref{FIG: La7Pd3 Magnetization}(a).  A rectangular sample of La$_{7}$Pd$_{3}$ was cut from the sample button to give a well-defined shape with a demagnetization factor $N = 0.13$~\cite{Aharoni1998}. A sharp change in the susceptibility marks the onset of superconductivity in La$_{7}$Pd$_{3}$ at $T^{\mathrm{onset}}_{\mathrm{c}} = 1.46(5)$~K in excellent agreement with a previous report~\cite{Pedrazzini00}. Between 0.5 and 1.1~K a full Meissner fraction ($\chi_{\mathrm{dc}} = -1$) is clearly visible indicating bulk superconductivity in La$_{7}$Pd$_{3}$. Several magnetization versus field loops were collected at different temperatures from 0.5 to 1.5~K in fields up to 10~mT. The lower critical field is estimated by measuring the field at which the flux first enters the sample, the first deviation from linearity in the magnetization versus applied field~\cite{Umezawa88}. Figure.~\ref{FIG: La7Pd3 Magnetization}(b) shows the lower critical field values extracted from these magnetization versus field loops. The temperature dependence of $H_{\mathrm{c1}}\left(T\right)$ can be described by the Ginzburg-Landau (GL) formula $H_{\mathrm{c1}}\left(T\right) = H_{\mathrm{c1}}\left(0\right)\left[1 - t_{\mathrm{r}}^{2}\right]$, where  $t_{\mathrm{r}} = T/T_{\mathrm{c}}$, giving $\mu_{0}H_{\mathrm{c1}}\left(0\right) = 6.9(2)$~mT. The magnetic susceptibility in the normal state shown in Fig.~\ref{FIG: La7Pd3 Resistivity} is almost temperature independent between 300 and 5~K with a shallow minimum at $\sim 50$~K. An upturn in $\chi\left(T\right)$ below 5~K is consistent with the presence of a small quantity of paramagnetic impurities (other rare-earths with sizable localized magnetic moments) present in the La used to prepare the samples. The normal state behavior of $\chi\left(T\right)$ is qualitatively similar to the only previous report~\cite{Pedrazzini00}. 

The temperature dependence of the electronic resistivity in La$_{7}$Pd$_{3}$ is shown in Fig.~\ref{FIG: La7Pd3 Resistivity}. The shape of $\rho\left(T\right)$ is characteristic of other NCS superconductors in the Th$_7X_3$ series~\cite{Pedrazzini00} as well as many intermetallic materials~\cite{Mayoh17, Joshi}. The resistivity at 300~K is 201.1(5)~$\mu\Omega$~cm and the residual resistivity $\rho_{0}$ at 2 K, just above $T_{\mathrm{c}}$, is $48.4(3)~\mu\Omega$ giving a residual resistivity ratio of $\sim4.2$ indicating poor metallic behavior. Using a simple Drude model~\cite{Ashcroft76, Mayoh17} we estimate the mean free path, $\ell= 12.4(1)$~nm.

\subsection{Specific heat measurements}
\label{Specific Heat}

\begin{figure}[p]
\centering
\includegraphics[width=0.8\columnwidth]{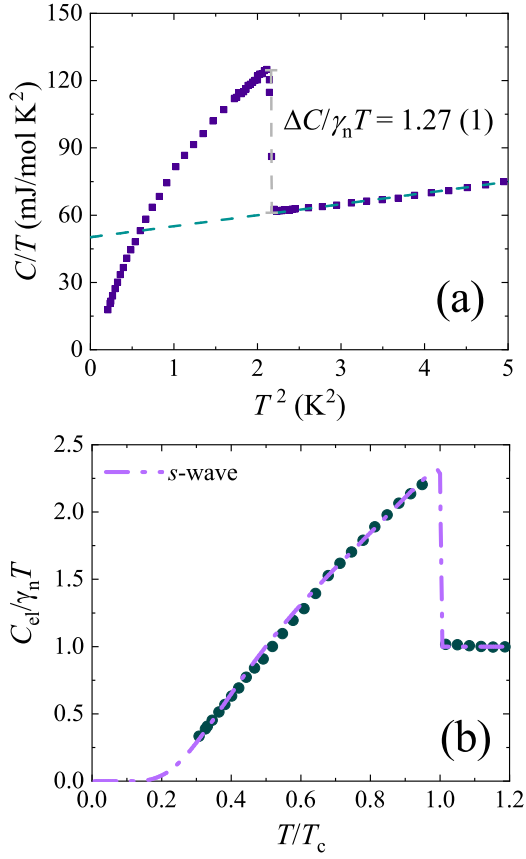}
\caption{(a) Temperature dependence of the zero-field heat capacity for La$_{7}$Pd$_{3}$ between 0.45 and 2.75~K showing a superconducting transition at $T_{\mathrm{c}} = 1.45(5)$~K. The shape of the $C$ versus $T$ is indicative of a typical type-II superconductor. Fitting the data above $T_{\mathrm{c}}$ in the normal-state using Eq.~(\ref{EQ: Sommerfeld}) gives $\gamma_{\mathrm{n}} = 50.2(2)$~mJ/mol~K$^{2}$. (b) Normalized electronic heat capacity $C_{\mathrm{el}}/\gamma_{\mathrm{n}}T$ versus the reduced temperature $T/T_{\mathrm{c}}$ in zero applied field. The dotted line shows a fit to the data for an isotropic $s$-wave gap made using Eqs.~\ref{EQ: Entropy1} and \ref{EQ: Entropy2}.}
\label{FIG: La7Pd3 Heat Capacity}
\end{figure}

The temperature dependence of the heat capacity in La$_{7}$Pd$_{3}$ between 0.45 and 2.75~K is shown in Fig.~\ref{FIG: La7Pd3 Heat Capacity}(a). A sharp jump in the heat capacity, $\Delta C$, is seen at $1.45(5)$~K indicating the onset of bulk superconductivity in La$_{7}$Pd$_{3}$. The superconducting transition in La$_{7}$Pd$_{3}$ is typical of that seen in a type-II superconductor. The normal-state heat capacity at $T> 1.45$~K can be modeled using 
\begin{equation} \label{EQ: Sommerfeld}
C\left(T\right)/T = \gamma_{\mathrm{n}}+\beta_{3}T^{2}+\beta_{5}T^{4},
\end{equation} 

\noindent to give the Sommerfeld coefficient ${\gamma_{\mathrm{n}} = 50.2(2)~\mathrm{mJ/mol~K^{2}}}$ and ${\beta_{3} = 4.86(4)~\mathrm{mJ/mol~K}^{4}}$. Writing ${\theta_{\mathrm{D}}=\left(\frac{12\pi^4}{5}Nk_{\mathrm{B}}/\beta\right)^{\frac{1}{3}}}$ gives the Debye temperature ${\Theta_{\mathrm{D}} = 159(2)~\mathrm{K}}$. The large $\gamma_{\mathrm{n}}$ is quite unusual for noncentrosymmetric superconductors, with the value for La$_{7}$Pd$_{3}$ surpassing some heavy-fermion superconductors such as CeCoGe$_{3}$ ($\gamma_{\mathrm{n}} = 32$~mJ/mol~K$^{2}$)~\cite{Settai07}. This suggests that there is an enhanced density of states at the Fermi level for La$_{7}$Pd$_{3}$. This value of $\gamma_{\mathrm{n}}$ is consistent with that seen in other La$_{7}X_{3}$~\cite{Pedrazzini00, Li17, Singh20, Nakamura17} compounds suggesting that this enhancement is a common feature of this group of superconductors. 

The electronic heat capacity in zero applied field in the superconducting state can be used to look for evidence for an unusual superconducting order parameter. The normalized entropy ${S_{\mathrm{el}}/\gamma_{\mathrm{n}} T_{\mathrm{c}}}$ is written as
\begin{equation}\label{EQ: Entropy1}
\dfrac{S_{\mathrm{el}}\left(T\right)}{\gamma_{\mathrm{n}} T_{\mathrm{c}}}=-\dfrac{6}{\pi^2}\dfrac{\Delta\left(0\right)}{k_{\mathrm{B}}T_{\mathrm{c}}}\int_{0}^{\infty}\left[f\mathrm{ln}(f)+\left(1-f\right)\mathrm{ln}\left(1-f\right)\right]d\epsilon,
\end{equation}
where $f\left(E\right) = \left[1+\exp\left(E/k_{\mathrm{B}}T\right)\right]^{-1}$ is the Fermi-Dirac distribution function with energy ${E=\sqrt{\epsilon^2+\left[\Delta\left(T\right)\right]^{2}}}$. $\epsilon$ is the energy of the normal state electrons and $\Delta\left(T\right)=\Delta\left(0\right)\delta\left(T\right)$ where $\Delta\left(0\right)$ is the magnitude of the superconducting gap at zero kelvin and the temperature dependence of the energy gap is approximated using a single-gap BCS expression~\cite{Carrington}, $\delta\left(T\right) = \tanh\lbrace 1.82\left[1.018\left(T_{\mathrm{c}}/T-1\right)\right]^{0.51}\rbrace$. The specific heat in the superconducting state is then calculated from~\cite{Johnston_2013}
\begin{equation}\label{EQ: Entropy2}
\dfrac{C_{\mathrm{el}}\left(T\right)}{\gamma_{\mathrm{n}} T_{\mathrm{c}}}=T \dfrac{d(S_{\mathrm{el}}/\gamma_{\mathrm{n}} T_{\mathrm{c}})}{dT}.
\end{equation}
\noindent Figure~\ref{FIG: La7Pd3 Heat Capacity}(b) shows that the normalized electronic heat capacity $C_{\mathrm{el}}/\gamma_{\mathrm{n}}T$ as function of the reduced temperature $T/T_{\mathrm{c}}$. The data and the fit are in good agreement indicating that the material has an isotropic, largely $s$-wave superconducting gap. $\Delta C /\gamma_{\mathrm{n}}T_{\mathrm{c}}= 1.27\left(1\right)$ and $\Delta\left(0\right)/k_{\mathrm{B}}T_{\mathrm{c}} = 1.67(1)$ are both slightly lower than the 1.43 and 1.76, respectively, expected from the BCS theory in the weak-coupling limit.  This may indicate a slightly diminished electron-phonon coupling strength in La$_{7}$Pd$_{3}$, however, see Section~\ref{MuSR}. 

\subsection{Upper critical field calculations}
\label{Hc2}
\begin{figure*}[p]
\centering
\includegraphics[width=2\columnwidth]{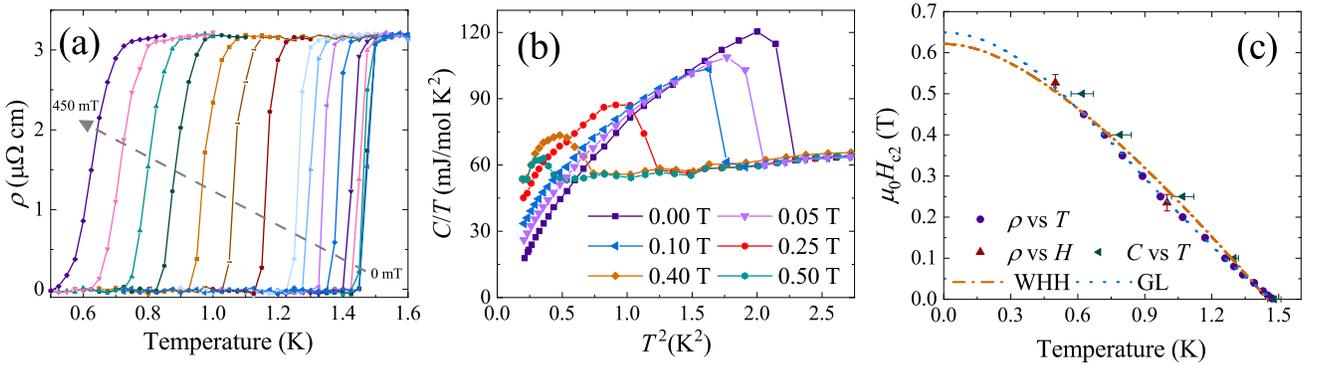}
\caption{(a) Temperature dependence of the resistivity of La$_{7}$Pd$_{3}$ showing the suppression and broadening of the resistive superconducting transition in applied fields from 0 to 450~mT. (b) $C/T$ versus $T^{2}$ for La$_{7}$Pd$_{3}$ showing the suppression and broadening of the superconducting transition as the applied field is increased from 0 to 500~mT. (c) Temperature dependence of the upper critical field for La$_{7}$Pd$_{3}$. The $H_{\mathrm{c2}}\left(T\right)$ values were extracted from the midpoints of the anomalies in $C\left(T\right)/T$ and the midpoints of the resistive transitions. The dotted and dash-dotted lines show fits to the $\mu_{0}H_{\mathrm{c2}}(T)$ data using the GL and WHH models~\cite{WHH, Mayoh17}, respectively.}
\label{FIG: La7Pd3 UpperCrit}
\end{figure*}

Fig.~\ref{FIG: La7Pd3 UpperCrit}(a) shows measurements of the resistivity as a function of temperature, $\rho\left(T\right)$, for La$_{7}$Pd$_{3}$ in various magnetic fields. There is a sharp superconducting transition at $T_{\mathrm{c}}= 1.47(5)$~K with a width of $\Delta T = 0.05$~K in zero-applied field that is suppressed with increasing magnetic field. The transition also broadens, $\Delta T = 0.23(1)$~K at 450~mT. Measurements of resistivity as a function of magnetic field at fixed temperature were also performed. The $T_{\mathrm{c}}$ values determined from these data are included in Fig.~\ref{FIG: La7Pd3 UpperCrit}(c). The temperature dependence of the heat capacity around the superconducting transition in different applied magnet fields is shown in Fig.~\ref{FIG: La7Pd3 UpperCrit}(b). Again, the transition is suppressed and broadened with increasing magnetic field.

Values for the upper critical field, $H_{\mathrm{c2}}$, of La$_{7}$Pd$_{3}$ were estimated from the midpoint of the superconducting transition in the resistivity data [see Fig.~\ref{FIG: La7Pd3 UpperCrit}(a)] and the heat capacity data [see Fig.~\ref{FIG: La7Pd3 UpperCrit}(b)]. Figure~\ref{FIG: La7Pd3 UpperCrit}(c) shows the upper critical field values as a function of temperature. The $H_{\mathrm{c2}}$ values exhibit a positive curvature close to $T_{\mathrm{c}}$. This behavior can be captured by the Ginzburg-Landau phenomenological model for upper critical fields, $H_{\mathrm{c2}}\left(T\right)=H_{\mathrm{c2}}\left(0\right)\left[\left(1-t_{\mathrm{r}}^{2}\right)/\left(1+t_{\mathrm{r}}^{2}\right)\right]$, giving a good fit to the data and an estimated upper critical field $\mu_{0}H_{\mathrm{c2}}\left(0\right) = 652(5)$~mT, almost 2.5 times the previously reported value~\cite{Pedrazzini00}, but still well below the Pauli limit which is similar to other superconductors in this series. Using  $H_{\mathrm{c2}}\left(0\right)=\frac{\Phi}{2\pi\xi^2}$ gives $\xi=22.5\left(1\right)$~nm.

La$_{7}$Pd$_{3}$ contains lanthanum and palladium and so it is anticipated that there may be a significant spin-orbit coupling contribution to the physics of the superconducting state of La$_{7}$Pd$_{3}$. The Werthamer-Helfand-Hohenberg (WHH) model allows for the inclusion of spin-orbit coupling in the upper critical field calculations~\cite{WHH, Mayoh17}. A fit to the $H_{\mathrm{c2}}\left(T\right)$ data was attempted using the WHH model as shown by the orange dash-dotted line in Fig.~\ref{FIG: La7Pd3 UpperCrit}(c) giving a slightly smaller $\mu_{0}H_{\mathrm{c2}}\left(0\right) = 620(3)$~mT. However, this model is unable to capture the curvature of the upper critical field values, cf. La$_{7}$Ir$_{3}$ where a WHH model provided a reasonable fit to the $H_{\mathrm{c2}}\left(T\right)$ data~\cite{Li17}.  

\section{$\mu$SR measurements, the superconducting order-parameter, and time-reversal symmetry breaking}
\label{MuSR}
The macroscopic superconducting state of La$_{7}$Pd$_{3}$ was probed using magnetization, resistivity, and heat capacity, however, in superconductors the microscopic magnetic environment formed by the vortex lattice can provide an essential insight into the superconducting state. Positive muons are an excellent probe of the local magnetic environment when implanted into a superconductor. The superconducting state of La$_{7}$Pd$_{3}$ has been investigated using transverse-field, longitudinal-field, and zero-field $\mu$SR. 

\begin{figure}[p]
\centering
\includegraphics[width=\columnwidth]{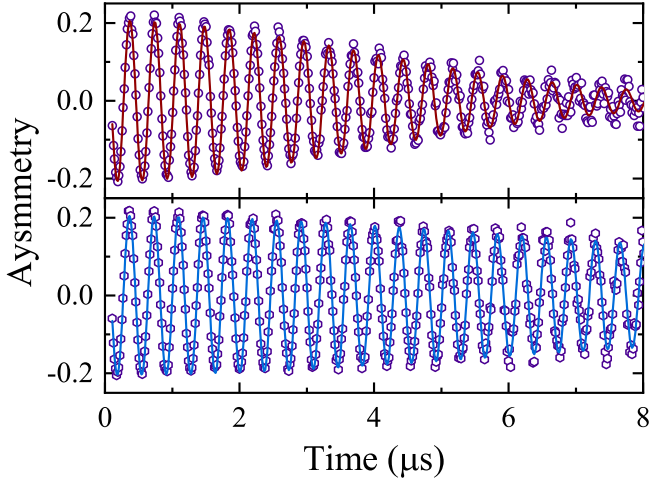}
\caption{Transverse-field $\mu$SR spectra for La$_{7}$Pd$_{3}$ collected at 100 mK (top) and 2.25 K (bottom) in an applied magnetic field of 20 mT. The solid lines are fits to data using Eq.~(\ref{EQ: Muon GaussRelax}). Below the superconducting transition temperature the field distribution of the FLL causes the spectra to be significantly depolarized. Above the superconducting transition temperature the randomly oriented array of nuclear magnetic moments continue to depolarize the muons but at a reduced rate.}
\label{FIG: La7Pd3 TFspectra}
\end{figure}

\begin{figure*}[p]
\centering
\includegraphics[width=2\columnwidth]{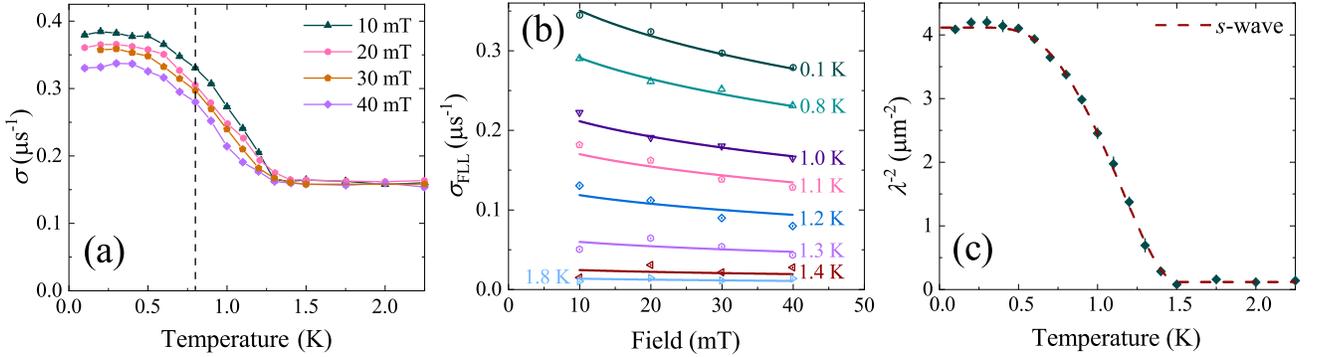}
\caption{(a) Temperature dependence of the total spin depolarization, $\sigma$, for La$_{7}$Pd$_{3}$ collected in fields between 10 and 50~mT. Isothermal cuts (the dashed line shows the cut made at 0.8~K) were used to calculate the field dependence of $\sigma_{\mathrm{FLL}}$ in La$_{7}$Pd$_{3}$. (b) Field dependence of the muon spin relaxation due to the flux line lattice, $\sigma_{\mathrm{FLL}}$, at different temperatures. The solid lines are fits to the data using Eq.~(\ref{EQ: Brandt Muon}). (c) Temperature dependence of the inverse square of the penetration depth, $\lambda^{-2}$. The dashed line is a fit to the data using Eq.~(\ref{EQ: Superfluid Density}).}
\label{FIG: La7Pd3 TFmuon}
\end{figure*}

Transverse-field spectra were collected at temperatures between 0.1 to 2.75~K in applied fields ranging from 10 to 50~mT, in the mixed state [$\mu_{0}H_{\mathrm{c1}}\left(0\right) = 6.9(2)$~mT]. In order to produce a well-ordered  flux line lattice, the sample was field-cooled before collecting the data on warming. Figure~\ref{FIG: La7Pd3 TFspectra} shows typical examples of the asymmetry spectra in the superconducting (0.1~K) and normal state (2.75~K). In the superconducting state, the muons depolarize quite rapidly due to the effects of the flux line lattice. A small amount of depolarization is still visible in the normal-state due to nuclear magnetic moments present in the sample. The oscillatory muon spectra can be fitted using a Gaussian function coupled with a cosinusoidal term for the muons implanted in the sample, and a simple cosinusoidal term for the muons implanted in the silver sample plate:
\begin{multline} \label{EQ: Muon GaussRelax}
G_{\mathrm{TF}}\left(t\right) = A_{1}\exp\left(-\frac{\sigma^{2}t^{2}}{2}\right)\cos\left(\gamma_{\mu}B_{1}t+\phi\right)\\
 + A_{2}\cos\left(\gamma_{\mu}B_{2}t+\phi\right).
\end{multline}
\noindent $A_{1}$ and $A_{2}$ are the sample and silver sample holder asymmetries, $B_{1}$ and $B_{2}$ are the average fields in the superconductor and silver plate, $\phi$ is the shared phase offset, $\gamma_{\mathrm{\mu}}/2\pi = 133.5$~MHz~T$^{-1}$ is the muon gyromagnetic ratio and $\sigma$ is the total depolarization rate. By fitting the spectra collected at different temperatures and fields using Eq.~(\ref{EQ: Muon GaussRelax}), the temperature dependence of $\sigma$ can be determined as shown in Fig.~\ref{FIG: La7Pd3 TFmuon}(a). The total depolarization rate, $\sigma$, is related to the depolarization due to the flux line lattice, $\sigma_{\mathrm{FLL}}$, and the nuclear moments, $\sigma_{\mathrm{N}}$, by
\begin{equation} \label{EQ: Muon depolarization}
\sigma^{2} = \sigma_{\mathrm{FLL}}^{2} + \sigma_{\mathrm{N}}^{2}.
\end{equation}
\noindent The nuclear depolarization rate is found to remain constant over all temperatures at $\sigma_{\mathrm{N}} = 0.162(1)~\mu\mathrm{s}^{-1}$. Since the upper critical field is comparable with the applied fields used in these measurements $\sigma_{\mathrm{FLL}}$ has a considerable field dependence. This is due to significant shrinking of the inter vortex distances within the flux line lattice as the applied field is increased. The effect of the vortex cores and the expected field dependence of the second moment of the field distribution have been calculated using different models. From calculations based on the Ginzburg-Landau model the field dependence of $\sigma_{\mathrm{FLL}}$ can be described using
\begin{multline}\label{EQ: Brandt Muon}
\sigma_{\mathrm{FLL}}\left[\mathrm{\mu s}^{-1}\right] = 4.854 \times 10^{4}\left(1-h_{\mathrm{r}}\right)\\
 \times\lbrace1+1.21\left(1-\sqrt{h_{\mathrm{r}}}\right)^{3}\rbrace\lambda^{-2}\left[\mathrm{nm}^{2}\right]
\end{multline}
where $h_{\mathrm{r}} = H/H_{\mathrm{c2}}$ is the reduced field and $\lambda^{-2}$ is the inverse square of the penetration depth~\cite{Brandt03}. By taking isothermal cuts of the data shown in Fig.~\ref{FIG: La7Pd3 TFmuon}(a) as denoted by the dashed line, Eq.~(\ref{EQ: Brandt Muon}) can be used to fit to the data, as shown in Fig.~\ref{FIG: La7Pd3 TFmuon}(b), and the penetration depth can be extracted. The resulting temperature dependence of $\lambda^{-2}$, which reflects the variation in the superfluid density, is shown in Fig.~\ref{FIG: La7Pd3 TFmuon}(b) and this can be used to investigate the nature of the superconducting gap in La$_{7}$Pd$_{3}$. In the clean limit, the magnetic penetration depth can be modeled using
\begin{equation}\label{EQ: Superfluid Density}
\frac{\lambda^{-2}\left(T\right)}{\lambda^{-2}\left(0\right)} = 1 + 2 \int_{\Delta\left(T\right)}^{\infty} \left(\frac{\partial f}{\partial E}\right)\frac{EdE}{\sqrt{E^{2}-\Delta^{2}\left(T\right)}},
\end{equation}
\noindent where $f$ is the Fermi-Dirac distribution function and the temperature dependence of the gap for an isotropic $s$-wave model is $\Delta\left(T\right) = \Delta\left(0\right)\delta\left(T\right)$ as in Section~\ref{Specific Heat}. The fit produced by this model is shown by the dashed line in Fig~\ref{FIG: La7Pd3 TFmuon}(c). The penetration depth at zero kelvin was calculated to be $\lambda(0) = 495(4)$~nm. The value of $\Delta\left(0\right) = 0.30(4)$~meV obtained gives $\Delta\left(0\right)/k_{\mathrm{B}}T_{\mathrm{c}} = 2.40(13)$ which is above both the BCS value of 1.76 and the value determined from the heat capacity measurements. Differences in $\Delta\left(0\right)/k_{\mathrm{B}}T_{\mathrm{c}}$ determined from heat capacity and $\mu$SR data have been observed in other superconductors, e.g.~\cite{muon-ReW}. Possible reasons for the difference include multigap superconductivity, a gap anisotropy, or a nodal gap due to a small triplet component leading to a reduced anomaly in heat capacity at $T_{\mathrm{c}}$~\cite{Golubov02, Weng2016, Mackenzie03}.

\begin{figure*}[p]
\centering
\includegraphics[width=2\columnwidth]{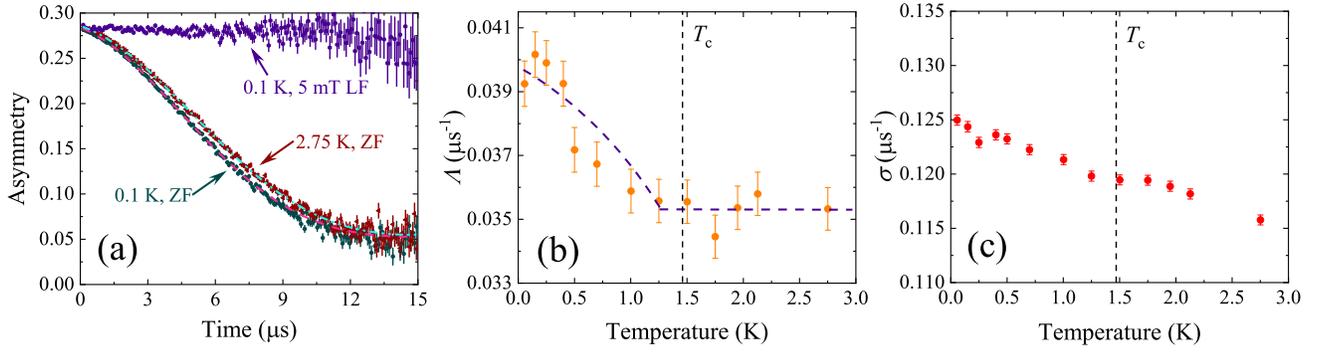}
\caption{(a) ZF and LF-$\mu$SR spectra collected at 0.1 (green) and 2.75~K (red), the data is fit using the Gaussian Kubo-Toyabe model (dashed lines). (b) Temperature dependence of the electronic relaxation rate $\Lambda$ can be seen to increase below 1.2~K just below $T_{\mathrm{c}}$. (c) Temperature dependence of the nuclear relaxation rate $\sigma$ shows no change at $T_{\mathrm{c}}$.}
\label{FIG: La7Pd3 LFmuon}
\end{figure*}

Zero-field measurements were performed on La$_{7}$Pd$_{3}$ to look for evidence of time-reversal symmetry breaking in the superconducting state. Examples of the asymmetry spectra collected at temperatures above (2.75~K) and below (0.1~K) the superconducting transition are shown in Fig.~\ref{FIG: La7Pd3 LFmuon}(a). These spectra exhibit considerable relaxation. The absence of any oscillatory component in the signals rules out the possibility of there being magnetic ordering in the sample. This observation is supported by measurements of the temperature dependence of the magnetic susceptibility at magnetic fields above $H_{\mathrm{c2}}\left(0\right)$ (see inset in Fig.~\ref{FIG: La7Pd3 Resistivity}). It can be assumed that the majority of the relaxation arises from the presence of static, randomly orientated nuclear moments, while the increased relaxation rate below $T_{\mathrm{c}}$ indicates the presence of additional small internal magnetic fields in the superconducting state. These small magnetic fields are associated with the onset of time-reversal symmetry breaking. To eliminate any possibility of the relaxation coming from spin fluctuations a small longitudinal field (LF) of 5~mT was applied, as shown in Fig.~\ref{FIG: La7Pd3 LFmuon}(a). The complete decoupling of the muons from the proposed relaxation channel in this small LF indicates that the spontaneous magnetic fields are static or at least quasi-static over the lifetime of the muon.

The response of the muons to the nuclear moments can be captured using the Kubo-Toyabe expression
\begin{equation} \label{EQ: Gaussian Kubo-Toyabe a}
G_{\mathrm{KT}}\left(t\right) = \frac{1}{3} + \frac{2}{3}\left(1-\sigma^{2}_{\mathrm{ZF}}t^{2}\right)\exp\left(-\frac{\sigma^{2}_{\mathrm{ZF}}t^{2}}{2}\right),
\end{equation}
where $\sigma_{\mathrm{ZF}}$ measures the width of the nuclear dipolar field experienced by the muons. The asymmetry can then be modeled by
\begin{equation} \label{EQ: Gaussian Kubo-Toyabe b}
G\left(t\right) = A_{0}G_{\mathrm{KT}}\left(t\right)\exp\left(-\Lambda t\right) + A_{\mathrm{bg}},
\end{equation}
where $A_{0}$ and $A_{\mathrm{bg}}$ are the sample and background asymmetries, respectively, and $\Lambda$ measures the electronic relaxation rate. The sample and background asymmetries were found to be constant at all temperatures. $\sigma_{\mathrm{ZF}}$ was found to decrease linearly with increasing temperature from 0.1 to 2.75~K across $T_{\mathrm{c}}$ [see Fig.~\ref{FIG: La7Pd3 LFmuon}(c)] while $\Lambda$ was found to be temperature independent above the superconducting transition and to increase immediately below $T_{\mathrm{c}}$ at $\sim1.2$~K [see Fig.~\ref{FIG: La7Pd3 LFmuon}(b)]. 

The increase in $\Lambda$ occurs at a temperature slightly below the superconducting transition temperature as determined from magnetization, resistivity, heat capacity, and TF-$\mu$SR measurements. This temperature difference between the signal marking the onset of TRS breaking and $T_{\mathrm{c}}$ is also seen in both La$_{7}$Ir$_{3}$~\cite{La7Ir3} and La$_{7}$Rh$_{3}$~\cite{Singh20}, two other members of this group of materials, as well as other superconductors such as PrPt$_{4}$Ge$_{12}$~\cite{PrPt4Ge12}.

\section{Comparisons with other superconductors exhibiting TRSB}
 
Magnetization, heat capacity, resistivity, and $\mu$SR measurements reveal that La$_{7}$Pd$_{3}$ is type-II superconductor with a $T_{\mathrm{c}}=1.45(5)$~K. Some of the fundamental superconducting parameters determined from these measurements are summarized in Table~\ref{Tab: CharNumbers}. Heat capacity and transverse-field $\mu$SR measurements indicate that the superconducting order parameter in La$_{7}$Pd$_{3}$ is dominated by a BCS-like $s$-wave component. The temperature dependence of the upper critical field is well fitted by a Ginzburg-Landau model which provides further evidence of conventional superconducting behavior.

On the other hand, zero-field $\mu$SR measurements reveal an increase in $\Lambda\left(T\right)$ at low temperature of $0.005\left(1\right)~\mu$s$^{-1}$. This is taken as evidence for the onset of time-reversal symmetry breaking although this change in $\Lambda\left(T\right)$ is only visible below $\sim1.2$~K. Similar behavior is observed in La$_{7}$Ir$_{3}$ and La$_{7}$Rh$_{3}$ (see Table~\ref{TAB: Compounds summary table}). It is possible that for all three materials TRS is broken at a second transition just below $T_{\mathrm{c}}$. Such a scenario has been suggested for LaNiC$_{2}$ and LaNiGa$_{2}$~\cite{Weng2016}. However, to date, there are no indications of any additional transition or evidence for two gap superconductivity in the La$_{7}X_{3}$ series of superconductors.

An increase in $\Lambda\left(T\right)$ at $T_{\mathrm{c}}$ in Sr$_{2}$RuO$_{4}$~\cite{Sr2RuO4}, LaNiC$_{2}$~\cite{LaNiC2}, and SrPtAs~\cite{Biswas13} is also attributed to time-reversal symmetry breaking. The $\Delta\Lambda$ and behavior of $\sigma_{\mathrm{ZF}}\left(T\right)$ are given in Table~\ref{TAB: Compounds summary table} for comparison. In Sr$_{2}$RuO$_{4}$, TRSB is thought to arise due to a degeneracy in the superconducting phase brought about by non-zero spin and orbital moments. This in turn allows for the creation of spontaneous moments near to grain boundaries and impurities due to variations in the superconducting order parameter~\cite{Sr2RuO4, Choi89, PrOsSb}. In LaNiC$_{2}$, the signature of TRSB results from hyperfine fields made by nonunitary spin triplet pairs~\cite{LaNiC2}. LaNiC$_{2}$ and Sr$_{2}$RuO$_{4}$ have unconventional gap structures, while La$_{7}$Pd$_{3}$, La$_{7}$Rh$_{3}$, and La$_{7}$Ir$_{3}$ all appear to have BCS-like $s$-wave gaps. A spin-split Fermi surface can look conventional if the magnitude of the two superconducting order parameters are similar, and singlet-triplet pairing would be difficult to differentiate from conventional $s$-wave pairing if $\bm{d}\left(\bm{k}\right)$ is small. In all the noncentrosymmetric compounds studied to date, the signature of TRSB is relatively weak when compared to the change in $\Lambda\left(T\right)$ seen in the $p$-wave superconductor Sr$_{2}$RuO$_{4}$. Measurements on single crystals of La$_{7}X_{3}$, which should more readily reveal any anisotropy or nodes in the gap, are essential to clarify the TRSB mechanism. The superconducting and normal-state properties of single crystal La$_{7}$Ni$_{3}$ point to it being a conventional superconductor~\cite{Nakamura17}, although no $\mu$SR results have yet been published to confirm whether time-reversal symmetry breaking is present in La$_{7}$Ni$_{3}$. This is important as the effects of spin-orbit coupling should be more prominent in La$_{7}X_{3}$ materials containing the heavier elements Ir, Pd, and Rh. 

Similar challenges are faced by those investigating the properties of the Re-based $\alpha$-Mn NCS superconductors. TRSB is reported for several compounds in this series, again with $s$-wave BCS-like gap structures~\cite{Re6Zr, Khan, Mayoh17, Pang18, Matano16,Re6Hfa, Re6Hfb, Re6Hfc,Re6Ti,Re24Ti5,Shang18,Shang18a} and rather small changes in $\sigma_{\mathrm{ZF}}\left(T\right)$ (see Table~\ref{TAB: Compounds summary table}). The observation of time-reversal symmetry breaking in the centrosymmetric rhenium exhibiting type-II superconductivity is particularly interesting~\cite{Shang18} and may indicate that TRSB in these Re-based compounds is not related to the noncentrosymmetric structure, cf. LaNiC$_2$ and LaNiGa$_2$~\cite{LaNiC2, LaNiGa2}. Pristine rhenium is a type-I superconductor and is driven type-II by shear strain. It would be interesting to investigate whether the increase in defects that accompanies this strain plays any role in the TRSB, and whether other elemental superconductors, including in this context La, show any evidence for similar effects. Further studies of centrosymmetric La$_{7}X_{3}$ materials will also provide vital information on the role the crystallographic structure plays in time-reversal symmetry breaking. In particular, studies of the centrosymmetric La$_{7}$Ru$_{3}$ should indicate whether a lack of a center of inversion is necessary for time-reversal symmetry breaking in the La$_{7}X_{3}$ compounds.

\section{Concluding Remarks}

Now that time-reversal symmetry breaking has been observed in La$_{7}$Pd$_{3}$ it joins the small number of NCS superconductors that also show this phenomena.
Single crystals of these superconductors are required in order to distinguish parity mixing effects in the materials that appear to have predominantly $s$-wave BCS-like superconducting gap structures. Studies of elemental superconductors and centrosymmetric superconducting analogues are also needed to clarify how important properties such as the lack of an inversion center and inhomogeneities are in driving time-reversal symmetry breaking in NCS superconductors. 

\begin{acknowledgments}
We are thankful for the technical support provided by Ali Julian and Patrick Ruddy.  This work is supported by UKRI and STFC through the provision of beam time at the ISIS Neutron and Muon source, UK~
This work is funded by the EPSRC, United Kingdom, through grants EP/T005963/1 and EP/M028771/1. 
\end{acknowledgments}

\bibliography{La7Pd3_DM_References}

\newpage

\begin{table}[p]
\caption{Superconducting parameters for La$_{7}$Pd$_{3}$.} \label{Tab: CharNumbers}
\begin{center}
\begin{tabular}{l c c c}
 \hline\hline
 Parameter & Value &  Unit \\
 \hline
$\mu_{0}H_{\mathrm{c1}}\left(0\right)$ & 6.9(2) & mT  \\
$\mu_{0}H_{\mathrm{c2}}\left(0\right)$& 652(5) &  mT  \\
$\lambda(0)$& 495(4) & nm  \\
$\xi(0)$& 22.5(1) & nm  \\
$\kappa_{\mathrm{GL}}$& 22.0(3) & dim.  \\
 \hline\hline
\end{tabular}
\end{center}
\end{table}

\begin{table*}[p]\centering
\begin{tabular}{l c c c C C L P c c}
 \hline\hline
 Compound & $T_{\mathrm{c}}$~(K) & $\Delta \Lambda$~($\mu \mathrm{s}^{-1}$) & $\Delta \sigma$~($\mu \mathrm{s}^{-1}$) & Channel of TRSB & Behavior of secondary channel & Instrument & Gap structure & CS or NCS & Ref.\\
 \hline
 La$_{7}$Pd$_{3}$ & 1.46 & 0.005 & - & $\Lambda$ & Linear inc. & MuSR & $s$-wave & NCS & This work.\\
 La$_{7}$Ir$_{3}$ & 2.25 & 0.011 & - & $\Lambda$ & Linear inc. & MuSR & $s$-wave & NCS & \onlinecite{La7Ir3,Li17} \\
 La$_{7}$Rh$_{3}$ & 2.65 & 0.005 & - & $\Lambda$ & Linear inc. & MuSR & $s$-wave & NCS & \onlinecite{Singh20} \\
 \hline
 Re$_{6}$Zr & 6.75 & - & 0.008 & $\sigma$ & Constant & MuSR & $s$-wave & NCS & \onlinecite{Re6Zr, Khan, Mayoh17, Pang18, Matano16} \\
 Re$_{6}$Hf & 5.98 & - & 0.005 & $\sigma$ & Linear inc. & MuSR & $s$-wave & NCS & \onlinecite{Re6Hfa, Re6Hfb, Re6Hfc} \\
 Re$_{6}$Ti & 6 & - & 0.009 & $\sigma$ & Linear dec. & MuSR & $s$-wave & NCS & \onlinecite{Re6Ti} \\
 Re$_{24}$Ti$_{5}$ & 6 & 0.006 & - & $\Lambda$ & Constant & GPS & $s$-wave & NCS & \onlinecite{Re24Ti5,Shang18} \\
 Re$_{0.82}$Nb$_{0.18}$ & 8.8 & - & 0.022 & $\sigma$ & - & MuSR, GPS, LTF & $s$-wave & NCS & \onlinecite{Shang18a} \\
 Re & 2.7 & - & 0.01 & $\sigma$ & - & GPS, LTF & $s$-wave & CS & \onlinecite{Shang18a} \\
 \hline
 LaNiC$_{2}$ & 2.7 & 0.009 & - & $\Lambda$ & Constant & MuSR &  & NCS & \onlinecite{LaNiC2, Chen13, Bonalde11} \\
 LaNiGa$_{2}$ & 2.1 & 0.01 & - & $\Lambda$ & Constant & MuSR & & CS & \onlinecite{LaNiGa2} \\
 \hline
 Sr$_{2}$RuO$_{4}$ & 1.5 & 0.04 & -- & $\Lambda$ & -- & M15  & $p$-wave & CS & \onlinecite{Sr2RuO4} \\
 \hline\hline
\end{tabular}
\caption{Selected properties of some noncentrosymmetric and centrosymmetric superconductors in which time-reversal symmetry breaking has been observed. $\Delta \Lambda$ and $\Delta \sigma$ are taken from the data presented in the listed references along with the channel in which the signal indicating TRS breaking is observed and the muon spectrometer used. These instruments include MuSR at ISIS, UK, the General Purpose Surface-Muon (GPS) and the Low Temperature Facility (LTF) instruments based at PSI, Switzerland, and the M15 beam line at TRIUMF, Canada.} \label{TAB: Compounds summary table}
\end{table*}

\end{document}